\documentstyle[aps,prl,multicol,floats,epsf]{revtex}

\begin{document}
\draft
\wideabs{

\title{Transport Anomalies and the Role of Pseudogap 
in the ``60-K Phase" of YBa$_{2}$Cu$_{3}$O$_{7-\delta}$}

\author{Kouji Segawa and Yoichi Ando}
\address{Central Research Institute of Electric Power 
Industry, Komae, Tokyo 201-8511, Japan}

\date{\today}
\maketitle

\begin{abstract}
We report the result of our accurate measurements of the $a$- and 
$b$-axis resistivity, Hall coefficient, and the $a$-axis 
thermopower in untwinned YBa$_{2}$Cu$_{3}$O$_{y}$ single 
crystals in a wide range of doping.
It is found that both the $a$-axis resistivity and the 
Hall conductivity show anomalous dependences on the 
oxygen content $y$ in the ``60-K phase" 
below the pseudogap temperature $T^*$.
The complete data set enables us to narrow down the possible 
pictures of the 60-K phase, with which we discuss a peculiar 
role of the pseudogap in the charge transport.

\end{abstract}

%74.25.Fy Transport properties
%74.62.Dh Effects of crystal defects, doping and substitution
%74.20.Mn Nonconventional mechanisms 
%74.72.Bk Y-based cuprates
}
\narrowtext

YBa$_{2}$Cu$_{3}$O$_{y}$ (YBCO) is one of the most 
intensively studied systems among the high-$T_c$ cuprates. 
Although it is known that an increase in the oxygen 
content $y$ from 6 to 7 causes the hole doping 
into the CuO$_2$ planes and 
leads to superconductivity, the dependence of $T_c$ on $y$ 
is non-trivial and there is a plateau at 
$T_c \simeq$ 60 K with $y$ of around 6.7 (``60-K plateau"
or ``60-K phase").
The origin of this plateau remains controversial 
and two different explanations have been discussed: 
one \cite{Veal} is to consider (partial) oxygen ordering 
in the Cu-O chain layers and asserts that the average 
valence of the Cu ions (and thus the hole concentration $n$) 
in the CuO$_2$ planes is unchanged in a certain range of $y$; 
the other \cite{Tallon,Akoshima} assumes that $n$ is 
continuously changing and relates the plateau 
to the ``1/8 anomaly",
which is a suppression of $T_c$ at the hole doping of 1/8 
per Cu due to a charge-density-wave instability (in other 
words, charged stripe formation \cite{Tranquada}). 
Because of this controversy and also of the difficulty in 
determining the actual hole concentrations in the CuO$_2$ 
planes, the understanding of the normal and superconducting 
states of underdoped YBCO remains far from satisfactory,
which has been a source of extra complications in 
elucidating the high-$T_c$ mechanism.

The difficulty in clarifying the origin of the 60-K plateau 
lies partly in the essentially inhomogeneous nature of the 
oxygen distribution in the Cu-O chain layers \cite{McCormack}; 
note that for a given $y$ the actual hole doping can differ 
depending on the arrangement of the O atoms \cite{note}, and 
the O atoms in the Cu-O chains can rather easily rearrange at 
room temperature, which causes the room-temperature (RT) 
annealing effect \cite{Lavrov}.  
These facts often make it difficult to 
achieve reproducibility in the experiments, and therefore 
very careful experiments armed with high-quality 
samples are necessary for obtaining reliable results.
Also, when measuring the in-plane transport properties,
the use of untwinned crystals is indispensable for 
extracting the intrinsic behavior of the CuO$_2$ planes.
Previous transport studies of untwinned YBCO crystals 
(e.g. Refs. \cite{Ito,Takenaka}) did not provide enough 
information on the $y$ dependence in the 60-K plateau 
region to discuss its origin.

In this Letter, we report the result of our careful study of 
the $a$-axis resistivity $\rho_a$ of high-quality untwinned 
YBCO single crystals for a wide doping range, where we paid 
particular attention to determining the absolute 
magnitude of the resistivity to the accuracy of 5\%.
Since the Cu-O chains run along the $b$-axis, 
in the $\rho_a$ measurements the electric current 
flows only in the CuO$_2$ planes and thus $\rho_a$ is not 
complicated with the conductivity of the chains \cite{Gagnon}.
We also measure the $b$-axis resistivity $\rho_b$, 
Hall coefficient $R_H$, and the $a$-axis thermopower at 290 K, 
$S_a(290{\rm K})$, all of which help elucidating the transport 
anomalies in the 60-K phase as a function of $y$.
We found that $\rho_a$ becomes remarkably independent of $y$ 
in the 60-K phase when the pseudogap opens, indicating that 
one of the following two situations is realized:
(i) the hole concentration in the planes, $n$, is essentially 
unchanged, or (ii) a change in $n$ is compensated by a change 
in the scattering time.
Both possibilities bear intriguing implications on not only 
the peculiar electronic state in the 60-K phase but also the 
role of the pseudogap in the charge transport.

The YBCO single crystals are grown in Y$_2$O$_3$ crucibles 
by a conventional flux method \cite{Segawa}.
The high purity of our crystals can be inferred from the optimum 
zero-resistance $T_c$, which is as high as 93.4 K observed for 
$y$ = 6.95 (the transition width is less than 0.5 K).
Before detwinning, the crystals are annealed to be tuned 
to the targeted oxygen content (both the annealing atmosphere 
and temperature should be varied to best tune the oxygen content 
over a wide range).  The crystals reported here are in the 
range of $y=6.45-7.0$.
The crystals are always quenched at the 
end of the high-temperature annealing. 
Detwinning is performed at temperatures below 220$^{\circ}$C
under a uniaxial pressure of $\sim$0.1 GPa while monitoring the 
crystal surface with a polarized-light microscope. 
We only measure samples that are perfectly detwinned.
The exact oxygen content is determined by iodometry.

After the preparations are finished, 
the samples are left at room temperature for 
at least a week for the oxygen arrangement to equilibrate.
(We do observe the RT annealing effect 
in the time scale of a few days, which 
slightly reduces the resistivity and increases $T_c$.)
Note that during the RT annealing 
the total oxygen content $y$ do not change, but the 
oxygen atoms tend to order locally to form longer Cu-O chains.
To check for the impact of the order of annealing/detwinning,
we prepared both pre-detwinned (annealed after detwinning) 
and post-detwinned (reverse order) crystals for $y$=6.65, 
6.75, and 6.80; differences in $\rho_a$ and $R_H$ 
between the pre- and post-detwinned crystals are 
confirmed to be within the experimental error.
This indicates that the extent of the oxygen ordering is 
reproducible in our samples as long as 
the annealing conditions are unchanged.

\begin{figure}[t!]
\epsfxsize=0.8\columnwidth
\centerline{\epsffile{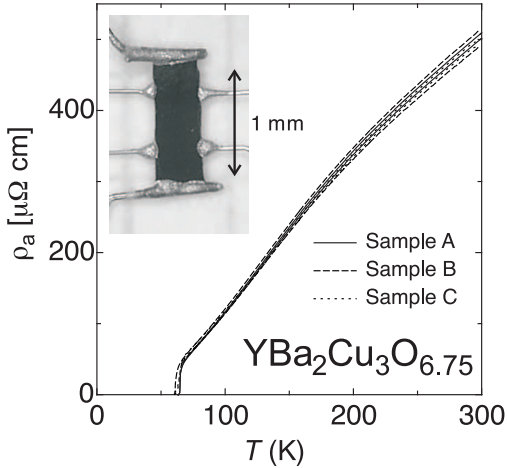}}
\vspace{0.2cm}
\caption{Six data sets of $\rho_a(T)$ measured on three samples 
at $y$=6.75, which represent the reproducibility and accuracy 
of our measurements.  Inset: Photograph of a typical sample with 
contacts.}
\label{fig1}
\end{figure}

Measurements of $\rho_a$ ($\rho_b$) are done with a standard 
ac four-probe method using samples that are at least two-times 
longer in the $a$ ($b$) direction.
(We note that it is easier to detwin crystals for 
$\rho_a$ measurements, for which the uniaxial pressure 
is applied to a smaller cross-sectional area.)
The Hall data are taken by sweeping the magnetic field to both 
plus and minus polarities at fixed temperatures. 
All the $R_H$ data shown here are measured with the 
electric current along the $a$-axis, and we confirmed that 
$R_H$ is essentially identical when the current is along the 
$b$-axis (i.e. Onsager's relation holds) \cite{SegawaHall}.
The thermopower is measured with a standard steady-state technique 
with a reversible temperature gradient of $\sim$1 K.

\begin{figure}[t!]
\epsfxsize=0.8\columnwidth
\centerline{\epsffile{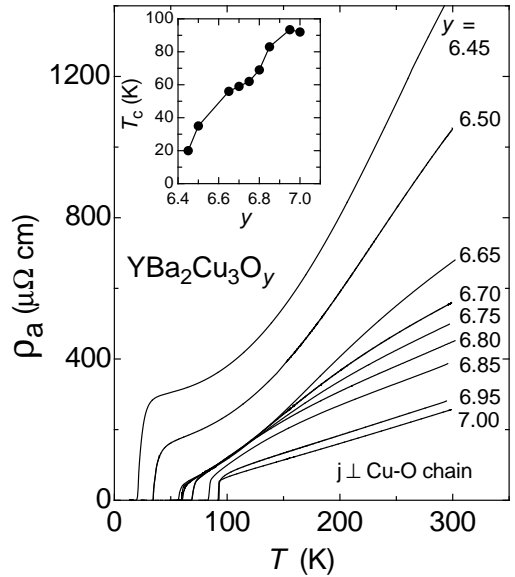}}
\vspace{0.2cm}
\caption{$T$ dependences of $\rho_a$ for untwinned YBCO 
crystals in 0 T.  
Inset: Phase diagram of zero-resistance $T_c$ vs $y$.}
\label{fig2}
\end{figure}

First we demonstrate the reproducibility and the accuracy 
of our measurement in Fig. 1, which shows the result of 
$\rho_a(T)$ measurements on three different samples, 
A, B, and C, all at $y$=6.75.  
As depicted in the inset to Fig. 1, the voltage is measured 
on the two sides of the crystals, so each sample yields two 
sets of $\rho_a(T)$ data; this is a good practice for 
checking the homogeneity of the sample.  
The thickness of the crystals is accurately determined 
by measuring the weight with 0.1 $\mu$g resolution. 
The largest source of error is the separation 
between the voltage contacts, which causes the uncertainty 
in the overall magnitude; use of a long sample can reduce 
this error to about 5\%, because our voltage contacts are 
defined by narrow gold pads whose width is $\sim$50 $\mu$m.  
Typically, our samples have the 
voltage-contact separation of $0.6-1$ mm, width of 0.3 mm, and 
the thickness of 60 $\mu$m.  As can be understood from Fig. 1, 
the reproducibility of the $\rho_a(T)$ data is 
very good and the scatter of the magnitude is consistent 
with the estimated error. 
$T_c$ is also very reproducible and its variation for a 
given $y$ is less than 5 K.
We note that $\rho_a(T)$ is measured on at least three 
crystals for each composition and the data 
are reproducible within the same order of accuracy as is 
demonstrated in Fig. 1. 
We pick up the data which sit in the middle of the spectrum 
to be representative of a given composition.

Figure 2 shows the temperature dependence of 
$\rho_a$ for the oxygen contents $y = 6.45-7.0$; this figure 
is a summary of the measurements of more than 30 samples.
The $y$=7.0 crystal ($T_c$=92.0 K) is slightly overdoped, 
while the $y$=6.95 crystal ($T_c$=93.4 K) is optimally doped and 
shows a strictly linear $T$ dependence with a negative intercept.
In the underdoped region, our samples show essentially similar 
behavior as was reported before \cite{Ito,Takenaka}; however,
what is remarkable here is that the $\rho_a(T)$ data for 
$y = 6.65-6.80$ show clear overlap below $\sim$130 K.
Note that in the underdoped YBCO the pseudogap opening 
can be inferred from a downward deviation from the 
high-temperature $T$-linear dependence below $T^*$ 
\cite{Ito}, and thus the data in Fig. 2 suggest that 
{\it the overlapping of $\rho_a(T)$ is observed in the 
pseudogapped state,} although $T^*$ inferred from the data 
is already above 300 K for $y$=6.65. 
Figure 3 shows the corresponding evolution of $\rho_b(T)$ 
with $y$.  Note that $\rho_b$ is generally smaller than 
$\rho_a$ for the same $y$, which is believed to be due to 
the finite chain conductivity \cite{Gagnon}.
The $\rho_b(T)$ data do not show as clear overlap in the 
60-K plateau region as the $\rho_a(T)$ data.

\begin{figure}[t!]
\epsfxsize=0.8\columnwidth
\centerline{\epsffile{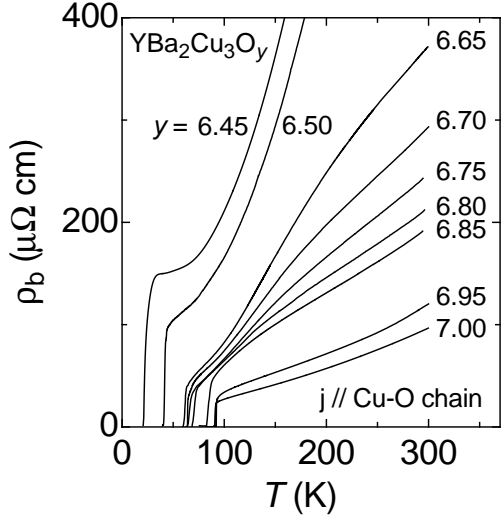}}
\vspace{0.2cm}
\caption{$T$ dependences of $\rho_b$ for untwinned YBCO 
crystals in 0 T.  Note the different scale compared to Fig. 2.}
\label{fig3}
\end{figure}

The Hall channel in the in-plane transport is also found to 
show an anomaly in the 60-K phase, which can be 
seen in the Hall conductivity $\sigma_{xy}$.
In YBCO, it is expected that the Cu-O chains contribute little 
to $\sigma_{xy}$ because of their one-dimensionality, 
so $\sigma_{xy}$ is governed by the properties of the planes.  
On the other hand, the Hall resistivity $\rho_{xy}$ is expressed 
as $\rho_{xy} \simeq \sigma_{xy}/(\sigma_{xx}\sigma_{yy})$ 
\cite{Ong} (where $\sigma_{xx} \simeq 1/\rho_a$ and 
$\sigma_{yy} \simeq 1/\rho_b$ in our case) and thus the 
Hall coefficient $R_H$ ($=\rho_{xy}/B$) reflects the 
properties of not only the planes but also the chains.
Therefore, $\sigma_{xy}$ is a better indicator of the 
properties of the planes compared to $R_H$ \cite{Uchida}.
The raw $R_H$ and the calculated 
$\sigma_{xy}$ [which is well approximated by 
$\rho_{xy}/(\rho_a\rho_b)$] 
are shown in Figs. 4(a) and 4(b), respectively, 
as functions of $y$ for 125 and 290 K. 
In Fig. 4(b), a non-monotonic $y$-dependence of $\sigma_{xy}$ 
is apparent at 125 K [where $\sigma_{xx} (\simeq 1/\rho_a$) 
is unusually $y$-independent], while the raw $R_H$ data 
[Fig. 4(a)] show relatively smooth change with $y$.
(Detailed account on the Hall data of our 
untwinned YBCO will be published elsewhere \cite{SegawaHall}.)
The nature of this anomaly in the Hall channel is probably 
best understood by the plot of the Hall mobility in the planes,
$\mu_H = \sigma_{xy}/(B\sigma_{xx})$ [Fig. 4(c)], 
which in principle does not include $n$.
One can clearly see that $\mu_H$ is anomalously enhanced 
near $y$=6.65, particularly at 125 K 
(in the pseudogapped state.)

It is generally believed that the room-temperature thermopower 
$S(290{\rm K})$ reflects the change in the hole concentration and 
thus may be used as a guide to estimate $n$ \cite{Tallon}. 
Figure 4(d) shows the $y$-dependence of the thermopower measured 
along the $a$-axis at 290 K, $S_a(290{\rm K})$.  
The $S_a(290{\rm K})$ data show a continuous change across the 
60-K phase, which is suggestive of $n$ changing with $y$; 
however, we cannot draw a definite 
conclusion from this, because there is a possibility that the 
density of states of the Cu-O chains gives some contribution 
to $S_a$.

\begin{figure}[t!]
\epsfxsize=0.95\columnwidth
\centerline{\epsffile{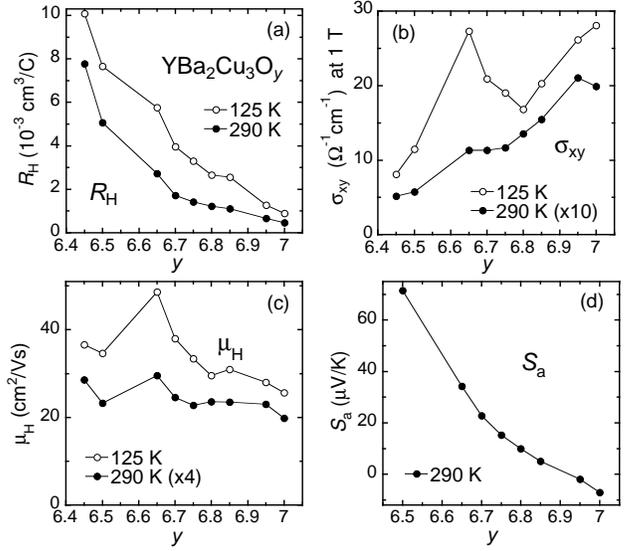}}
\vspace{0.2cm}
\caption{$y$ dependences of (a) raw $R_H$, (b) Hall conductivity 
$\sigma_{xy}$ (calculated for $B$=1 T), 
(c) Hall mobility $\mu_H$, at 125 and 290 K, and (d)
$a$-axis thermopower $S_a$ at 290 K.}
\label{fig4}
\end{figure}

Now let us discuss the implications of the above results.  
First, we note that the $y$-independence of $\rho_a$ (observed 
below $\sim$130 K in the 60-K phase) has 
two possible origins: (i) both the carrier concentration $n$ 
and the scattering time $\tau$ remain unchanged with $y$, or 
(ii) a change in $n$ is compensated by a change in $\tau$.  
Since it is difficult to conclusively identify the origin 
based on the data presented here, we should fully discuss 
the possible nature of the 60-K phase for the two cases.

If case (i) is true, we must understand 
why at high temperatures $\rho_a$ changes with $y$.  
In Fig. 2, it looks like 
the $\rho_a(T)$ data for $y = 6.65-6.80$ gradually converge 
to a single curve as the temperature is lowered below $T^*$; 
this trend can be interpreted to mean that 
there is some $y$-dependent scattering above $T^*$ that is 
gradually wiped out in the pseudogapped state.
Remember that the pseudogap is a partial gap near ($\pi$,0) and 
(0,$\pi$) on the Fermi surface \cite{PG}, and therefore any 
scattering that is concentrated near these regions 
of the Fermi surface is expected to be wiped out when the 
pseudogap opens.
Since the scattering caused by the AF fluctuation is 
concentrated in the ``hot spots" \cite{Stojkovic}, which 
correspond to the gapped regions in the pseudogapped state, 
the magnetic scattering could cause $\rho_a$ to be 
$y$ dependent above $T^*$ even when $\tau$ remains 
unchanged with $y$ below $T^*$.

If this is really the case, the strength of the magnetic 
fluctuations must be changing with $y$ in the 60-K phase
while $n$ stays unchanged;
we can construct an argument for this by recalling 
the role of the oxygen in the Cu-O chains. 
Within the oxygen ordering scenario \cite{Veal}, 
an addition of oxygen converts Cu$^{1+}$ on the chains 
into Cu$^{2+}$, adding spins onto the chain layers instead of 
adding holes into the CuO$_2$ planes in the 60-K phase. 
These Cu spins on the chains mediate the magnetic 
coupling between the bilayers \cite{TranquadaYBCO}, 
causing a suppression of the AF fluctuations in the 
CuO$_2$ planes.
We can thus argue for case (i) that the increase in $y$ may 
lead to weaker AF fluctuations, which cause the quasiparticles 
in the hot spots to be scattered less, 
leading to a decrease in $\rho_a(T)$ at high temperatures.
Note that the peak in $\mu_H$ at $y$=6.65 seems to be 
inconsistent with the assumption of $y$-independent $\tau$ 
below $T^*$, but this inconsistency might be resolved by 
considering a scattering-time separation \cite{Anderson,Coleman}.

Next, if case (ii) is true, 
the transport properties demonstrate an unusual variation of the 
scattering events in the 60-K phase. 
In this case, we can interpret that $\mu_H$ directly 
reflects the variation of $\tau$ with changing $n$; 
namely, we can infer that $\tau$ at 125 K is notably enhanced 
as $y$ is decreased from 6.8 to 6.65 in the 60-K phase.
If this enhancement in $\tau$ compensates a decrease in $n$, 
$\rho_a$ becomes $y$-independent as is observed in Fig. 2.
Therefore, in this scenario, the anomalous $y$-dependence of 
$\mu_H$ and the unusual overlap of $\rho_a(T)$ 
have a common physical origin.
The fact that the compensation is visible only below $T^*$ 
indicates that the anomalous enhancement of $\tau$ 
only takes place in the pseudogapped state, which suggests that 
this anomaly is of electronic origin.
This observation reveals a novel aspect of the pseudogap in YBCO, 
which might help elucidating the origin of the pseudogap itself.
We note that the $y$-dependence of $S_a(290{\rm K})$ seems to 
support this scenario, although it is not conclusive.

Within the picture of case (ii), the 60-K plateau in $T_c$ 
may be understood to be caused by the anomalous $y$-dependence 
of $\mu_H$ in this doping region, because $T_c$ is expected 
to be reduced when the carrier scattering is enhanced (which 
leads to smaller $\mu_H$).  Therefore, if case (ii) 
is true, the 60-K phase corresponds to a particular doping region 
where $\tau$ in the pseudogapped state is anomalously enhanced 
as $n$ is decreased, and the peculiar features in $\rho_a$, 
$\sigma_{xy}$, and $T_c$ can essentially be understood by the 
unusual doping dependence of $\tau$.
If, on the other hand, the oxygen ordering causes the case (i) 
to be realized, the 60-K phase of YBCO offers a unique case 
where the hole concentration is essentially unchanged 
while the strength of the AF fluctuations keeps changing 
with $y$; within this scenario, the plateau in $T_c$ suggests 
that the charge carriers are {\it protected} from the 
bare AF fluctuations once the pseudogap opens and 
the occurrence of superconductivity is governed by $n$. 
In both cases, our data bear fundamental implications on the 
electronic state in YBCO.

Lastly, we comment that our result does not 
exclude the possibility that the hole doping 
happens to be 1/8 per Cu in the planes for $y \simeq 6.7$.  
In view of the reports that suggest the existence of 
1/8-anomaly-like features in Ca-doped YBCO 
\cite{Tallon,Akoshima}, it is perhaps sensible to assume 
that both an unusual electronic state 
and the 1/8 doping are realized in the 60-K plateau region.

In summary,
we found a remarkable overlap of the $\rho_a(T)$ data of 
untwinned YBa$_{2}$Cu$_{3}$O$_{y}$ crystals 
in the 60-K phase below $\sim$130 K.  Moreover, the transport 
in the Hall channel is also found to show anomalous behavior 
in the same region.
We discuss that two different scenarios can potentially 
explain the observed anomalies, and in both scenarios the 
pseudogap appears to play a key role; in any case, it is clear 
that an unusual electronic state is responsible for the 
anomalies in the 60-K phase.
Further studies of this unusual region is clearly desirable, 
and the set of data presented here would serve as a 
touchstone to test any model for the 60-K phase.

We would like to acknowledge A. N. Lavrov, N. P. Ong, 
I. Tsukada, and S. Uchida for helpful discussions, 
and Y. Abe and Y. Hanaki for technical assistance.

% Place here the list of the references:
%
\medskip
\vfil


\begin{references}
\bibitem{note}
The Cu(1) atom in the chain layer becomes Cu$^{1+}$ 
when both of the neighboring O sites are vacant, while it becomes 
Cu$^{2+}$ when the O site(s) are filled \cite{Veal}.
\end{references}
\end{document}